\def\papertype{preprint}

\def\pSS{sissa}
\def\pMS{manuscript}
\def\pPP{preprint}

\def\figfont{\ifx\papertype\pPP \small\sf \fi}

\def\Mpc{{\rm\,Mpc}}
\def\msun{{\,M_\odot}}

\def\cm{{\rm\,cm}}

\def\gram{{\rm\,g}}
\def\gcm3{\gram/\cm^3}

\def\ie{{\it i.e.}~}

\def\cm{{\rm\,cm}}

\def\gram{{\rm\,g}}
\def\gcm3{\gram/\cm^3}

\newbox\grsign \setbox\grsign=\hbox{$>$} \newdimen\grdimen \grdimen=\ht\grsign
\newbox\simlessbox \newbox\simgreatbox
\setbox\simgreatbox=\hbox{\raise2.0pt\hbox{$>$}\llap
     {\lower3pt\hbox{$\sim$}}}\ht1=\grdimen\dp1=0pt
\setbox\simlessbox=\hbox{\raise2.0pt\hbox{$<$}\llap
     {\lower3pt\hbox{$\sim$}}}\ht2=\grdimen\dp2=0pt
\def\gta{\mathrel{\copy\simgreatbox}}
\def\lta{\mathrel{\copy\simlessbox}}

\def\h50{h_{50}}

\def\M9{\left({M\over 10^9\msun}\right)}

\def\m4{\dot{M}_4}

\def\v100{v_{100}}

\ifx\papertype\pMS
\documentstyle[epsf,12pt,aasms4,astrobib]{article}
\fi
\ifx\papertype\pPP
\documentstyle[epsf,12pt,aaspp4,astrobib]{article}
\fi
\ifx\papertype\pSS
\documentstyle[epsf,12pt,aaspp4,astrobib]{article}
\fi

\ifx\papertype\pPP
\makeatletter
\def\thebibliography{\section*{References\@mkboth
 {REFERENCES}{REFERENCES}}\list
 {[\arabic{enumi}]}{\leftmargin 1em\labelwidth\z@\labelsep\z@\itemindent -1em
 \parsep 0.0ex
 \usecounter{enumi}}
 \def\newblock{\hskip .11em plus .33em minus -.07em}
 \sloppy\clubpenalty4000\widowpenalty4000
 \sfcode`\.=1000\relax}
\makeatother
\fi

\begin{document}


\clearpage
\title{The Hubble Deep Field and the Disappearing Dwarf Galaxies}

\author{HENRY C. FERGUSON}
\affil{Space Telescope Science Institute, 3700 San Martin Drive\\
Baltimore, MD 21218\\
email: {ferguson@stsci.edu}
}
\author{ARIF BABUL}
\affil{Department of Physics \& Astronomy\\
University of Victoria\\
3800 Finnerty Road, Victoria, BC V8W 3P6\\
email: {babul@alMuhit.phys.uvic.ca}
}


\ifx\papertype\pPP \singlespace \fi
\ifx\papertype\pSS \singlespace \fi

\begin{abstract}

Several independent lines of reasoning, both theoretical and
observational, suggest that the very faint ($B\gta 24$) galaxies seen
in deep images of the sky are small low-mass galaxies that experienced
a short epoch of star formation at redshifts $0.5\lta z\lta 1$ and have since faded
into low luminosity, low surface brightness objects. Such a scenario,
which arises naturally if star formation in dwarf galaxies is delayed
by photoionisation due to the metagalactic UV radiation field, provides
an attractive way to reconcile the Einstein-de Sitter ($\Omega = 1;
\Lambda = 0$) cosmological model to the steeply rising galaxy counts
observed at blue wavelengths. Babul and Ferguson (1996) constructed a
specific realisation of this model, deriving the dwarf galaxy mass
function from the CDM power spectrum, and arguing that the gas
in these halos will recombine at $z \sim 1$.  The Hubble Deep Field
(HDF) images provide a stringent test of this model. We compare the
model to the data by constructing simulated images that reproduce the
spatial resolution and
noise properties of the real data and carrying out source detection and
photometry for the simulations in the same way they were carried out
for the real data. The selection biases and systematic errors
that are inevitable in dealing with faint galaxies are thus
incorporated directly into the model. We compare the model predictions
for the counts, sizes, and colours of galaxies observed in the HDF, and
to the predictions from a low $q_0$ pure-luminosity-evolution (PLE) model.
Both models fail to reproduce the observations.  The low $q_0$ model
predicts far more Lyman-break ``dropouts'' than are seen in the data.
The fading dwarf model 
predicts too many remnants: faded dwarf galaxies in the redshift range
$0.2 < z < 0.5$ that should be detectable in the HDF as low-surface
brightness red objects but are not seen.  {\it If fading dwarf galaxies
are to reconcile the Einstein-de Sitter geometry to the counts, then
the dwarf population must (a) form earlier than $z \sim 1$, with a
higher initial luminosity; (b) have an initial-mass function more
heavily weighted toward massive stars than the Salpeter IMF; or (c)
expand much more than assumed during the supernova wind phase.}

\end{abstract}

\ifx\papertype\pMS
\clearpage
\fi

\section{Introduction} \label{sec1}

\ifx\papertype\pPP \singlespace \fi
\ifx\papertype\pSS \singlespace \fi

Any self-consistent theory of galaxy evolution and cosmology must now
pass the test of matching the distribution of galaxy properties in the
Hubble Deep Field. The theory must simultaneously match the counts,
redshift and colour distributions, clustering properties, and size
distribution of faint galaxies.  Even at ground-based depths (e.g.
\citeNP{CFRS1,ECBHG96,CSHC96}), matching the observed
distributions within an Einstein-de Sitter ($\Omega=1$, $\Lambda=0$)
cosmology has required either introducing heuristic modifications to
the local luminosity function and the evolution of low-mass galaxies
\cite{PD95}, or introducing additional physical processes
into theories of galaxy evolution beyond simple passive evolution of
the stellar populations.  In hierarchical prescriptions (e.g.
\citeNP{KGW94,CAFNZ94}) the additional physics has been a
detailed treatment of the merging histories of galaxies, coupled with
prescriptions for how star formation proceeds during the merger events.
The primary physical process driving the counts in these models is
triggered star formation. An alternative view \cite{BR92,BF96}
is that the important physics driving the counts at very
faint magnitudes is the delayed formation of dwarf galaxies, where the
cooling and collapse timescales of the interstellar gas is governed by
photoionisation by the UV background.

The purpose of this paper is to test this latter model against
the Hubble Deep Field (HDF) observations. The comparison involves
constructing simulated HDF images from Monte-Carlo realisations
of the underlying model and then analyzing them  in exactly
the same manner as the actual images. In this way all of the
details of galaxy selection and photometry are properly taken into
account.

It is illustrative to consider the $q_0=0.5$ dwarf-dominated model not
in isolation, but in comparison to a low$-\Omega$ giant-dominated
model. \citeN{GK95} and  \citeN{PBZ96} have demonstrated that {\it for
low values of $q_0$} straightforward ``pure-luminosity evolution''
models, without large populations of dwarf galaxies or substantial
number-evolution, can be reasonably successful at matching the observed
counts and redshift distributions (to 1995 ground-based limits).  Thus,
as a foil to the dwarf-dominated $q_0 = 0.5$ model, we construct a
giant-dominated $q_0 = 0.01$ model and carry out the same comparisons
to the observations.

We emphasize that our goal in this paper is not to tune the models
to try to match the observations, but to test the models in their
simplest form to illustrate their differences (and their deficiencies)
at the depths probed by the Hubble Deep Field. The models
incorporate the simplest, most conservative, assumptions for the
luminosity function of local giant galaxies, their stellar initial mass
functions and evolution, their star-formation histories, and their
surface-brightness distributions.  The predictions of both models turn
out to be very sensitive to the redshift of formation, which has not
been tuned {\it post-facto} to try to meet the constraints imposed by
the observations. The dwarf galaxies are incorporated into the $q_0 =
0.5$ model with only three free parameters (described in section 1).
These parameters were fixed from physical arguments and ground-based
observations \cite{BF96} and have {\it not} been tuned to
match the HDF observations.

In \S 2, we briefly review the $q_0=0.5$ dwarf-dominated model 
and the $q_0=0.01$ model, and discuss the construction of simulated 
HDF images from the corresponding Monte-Carlo realisations.
In \S 3, we compare the results for the two models to the 
observations. In \S 4 we outline the implications of these
comparisons, specifically identifying changes that might be made
to each model to bring it more into agreement.

\section{The Models} \label{sec2}

\subsection{The $q_0 = 0.5$, bursting-dwarf model}
In a $q_0=0.5$ cosmology, matching the galaxy number counts and redshift
distributions observed to ground-based limits requires a large
population of low-luminosity galaxies at moderate redshifts.   
\citeN{BF96} outlined a dwarf-dominated model that appeared capable
of reproducing existing ground based observations.  The model consists of
``giant'' galaxies (types E through Sdm) that form at $z_f=5$ and a
large population of dwarf galaxies that begin forming stars
at $z \approx 1$. (Babul \& Ferguson refer to these objects as 
``boojums,'' for blue objects observed just undergoing a moderate starburst).
The comoving number density of each galaxy type is
conserved, but the dwarfs are entirely gaseous until $z < 1$.

In hierarchical clustering scenarios with realistic initial conditions
on galactic and sub-galactic scales (\ie spectral index $-2\lta n \lta
-1$), the distribution of $M\sim 10^8$--$10^9\msun$ halos is a steep
function of mass ($d\ln N/d\ln M \approx -2$) and the bulk of the halos
are expected to form at $z\gta 3$.  Under ``ordinary'' conditions, the
gas in these halos would rapidly cool, collect in the cores and undergo
star-formation.  Studies of Ly$\alpha$ clouds, however, suggest that
the universe at high redshifts is permeated by an intense metagalactic
UV flux.  This UV background will photoionise the gas and hence
prevent it from cooling and collecting in the cores of low mass halos
until $z\lta 2$ when the photoionising background begins to decline
rapidly \cite{BR92,QKE96}.  In our model, the power
spectrum of density fluctuations that grow into halos of circular
velocity less than $35 \rm\, km\,s^{-1}$ is described by the standard
cold dark matter (SCDM) power spectrum normalized such that the rms
density fluctuations on the scale of $8 h^{-1} \rm Mpc$ are
$\sigma_8=0.67$.  The comoving number density of minihalos existing at
$z\approx 1$, estimated using the analytic Press-Schecter
\citeyear{PS74} formalism, is
\begin{equation}\label{dwarflf} {dN\over d\ln(M)}\approx 2.3
\left({M\over 10^9\msun}\right)^{-0.9}\Mpc^{-3}.  \end{equation} The
mass function of dwarf galaxies is thus fixed with no free parameters.
Once the gas in dwarf galaxy halo can recombine and cool,
star-formation occurs rapidly, ceasing at $10^7$ years (the typical
lifetime of massive stars)  when supernova explosions heat and expel
the gas.  The probability of such a burst occuring in a dwarf galaxy is
assumed to decline exponentially with redshift from $z=1$, with the
timescale ($t^* = 2 \times 10^9$ yr) fixed by the requirement that the model 
must match the
observed B-band redshift distribution of \citeN{GECBAT95}.

Apart from the dwarf-galaxy mass function, which is fixed, 
the three most important parameters of the model are (1) star-formation efficiency
\footnote{Babul \& Ferguson (1996) incorrectly computed the total stellar mass 
for their dwarf galaxies; their assumed star-formation efficiency of 30\% resulted
in luminosities a factor of 2.8 too high, for the adopted IMF. 
To make the results directly comparable, we have kept the same total luminosities
for the dwarf galaxies, but lowered the corresponding star-formation efficiency 
to 10\% (for a Salpeter IMF from 0.1 to 100 $M_\odot$).
Nevertheless, 
as most of the light is produced by massive stars and most of the stellar
mass is in low-mass stars, the conversion from dwarf-galaxy mass to dwarf-galaxy
luminosity carries with it large uncertainties.}
, assumed to be 10\%, 
(2) $t^*$, the burst probability timescale, and (3) the redshift, 
$z_{f}$, at which the dwarfs start to form stars --- which
could plausibly be as high as $z=1.5$ and as low as $z=0$ and could vary as
a function of mass. The choice of these parameters is discussed by
\citeN{BF96}; they have not been adjusted to fit the HDF data.

The parameters for the giant galaxies in the model are similar to those
used in pure luminosity evolution (PLE) studies \cite{YT88,RG88,PBZ96}.
The luminosity functions for the giants (E through Sdm) are Gaussian,
tuned to approximate the type-dependent luminosity functions given
by \citeN{BST88annrev}. The bulge/disk total mass ratios and scatter are
tuned for each type to match the observations of \citeN{SdV86} for
local galaxies. This combination of parameters has been shown 
\cite{BF96} to provide a good fit to the overall luminosity function from the 
APM survey \cite{LPEM92}.

The parameters of the different distribution functions
are summarised in Table \ref{tabgalpars}, and described in more detail
in \citeN{BF96}.
Table \ref{tabgalpars} lists, for each galaxy type: $N_0$, the co-moving
space-density integrated over the entire luminosity function; the
characteristic absolute magnitude ($M_{BJ}$ for the Gaussian luminosity functions,
$M_{BJ}^*$ for the Schechter function, and $M_0$ for the power-law
mass function used for the dwarfs); the width of the Gaussian luminosity
function $\sigma$ or the power-law exponent $\alpha$ of the faint-end of the 
luminosity function; the mean bulge/total luminosity ratio in the
$B_J$ band and the Gaussian scatter $\sigma$ about that mean; the
redshift of formation $z_f$, or maximum redshift of formation $z_{max}$;
and the $e$-folding timescales of star-formation of bulge and disk.

Upon formation, all galaxies (giants and dwarfs) evolve only in
their stellar populations.
That is, there is no merging and there are no subsequent bursts of
star formation. The different colours of present-day giant galaxies are
reproduced by adjusting the star-formation timescale ($e$-folding time)
of the model. Galaxy disks form stars at roughly constant rate from
$z=5$ to $z=0$. Bulges form stars in a burst of duration $2 \times 10^8$
years.\footnote{The $e$-folding time of the initial star-formation episode
adopted in this paper is revised from that adopted by \citeN{BF96},
and is the average of the timescales for the onset of galactic winds
from the various models reviewed by Gibson (1996). }
\nocite{Gibson96}
Spectral evolution is computed using the Padova isochrones
\cite{BBCFN94} and stellar atmosphere models. The intrinsic
spectra of the galaxies are attenuated by the mean expected intergalactic
HI absorption using the models of \citeN{Madau95}.
This attenuation has a significant effect on the predicted numbers and
colours of galaxies with $z > 2.5$ \cite{MFDGSF96}. 

\subsection{The $q_0 = 0.01$, Pure Luminosity Evolution}
The $q_0=0.01$ model is identical to the
above model with two exceptions. 
First, there is much more volume at high redshift in the low $q_0$ model, and 
hence more high-redshift galaxies are predicted per unit area on the sky.
Second, the model does not include
the large population of boojums turning on at $z\approx 1$.
Instead the faint end of the luminosity function is populated by
galaxies with a star-formation timescale of $3 \times 10^{10} \rm \,yr$
and a \citeN{Schechter76} luminosity function with 
parameters $\alpha = -1.3$ and  $M^*_{B} = -16$, and an integrated 
co-moving density $n_0 = 0.2 \,\rm Mpc^{-3}$ from $3.5 \times 10^{-3}$ to $10
L^*$. In 
other words, the faint-end of the luminosity function is modeled
as a population of dwarf irregular galaxies with a luminosity function
slope intermediate to the flat value measured by \citeN{LPEM92} and 
\citeN{LKSLOTS96}
and the steeper values measured in the CfA redshift survey
\cite{MHG94} and in nearby clusters \cite{DePPHM95}.
It should be noted that similar models PLE \shortcite{PBZ96,YP94}, 
have been shown to provide a reasonable fit to the ground-based counts and redshift distributions,
provided $q_0$ is low. However, this may have been to a certain extent
fortuitous, as none of the previous models included intergalactic absorption, which
could in principle have a significant effect on the counts and colours
of faint galaxies \cite{Madau95}.

\subsection{Simulations and Galaxy Photometry}
Having set all the distribution functions and evolutionary parameters, a
Monte-Carlo procedure is used to produce a list of galaxy
redshifts, masses, ages, bulge $r_e$ and disk $\alpha$ (in kpc and arcsec), 
and magnitudes in various bands (for bulge and disk components, as well as 
the total) for each of the galaxies.  Magnitudes are computed by integrating 
the properly redshifted and $k$-broadened spectra for the bulge and disk
components through the filter bandpasses given in the HST WFPC-2 handbook
\cite{WFPC2handbook} as implemented in the IRAF synphot task.

These catalogs of simulated galaxies are used as input to the
IRAF artdata task to construct simulated images of the HDF. 
A noiseless image of the model galaxies in each band is constructed
and convolved with the HST point-spread function (measured from the
unsaturated blue star near the center of chip 4 in the real observations).
We have simulated only the wide field (WF) chips 
for the comparisons, adopting the plate scale (0.04 arcsec/pixel) of
the HDF ``drizzled'' images \cite{WBDDFFGGHKLLMPPAH96}.
A separate noise image is constructed
from a model that includes Poisson errors on the sky background and readout
noise for each of the exposures. This noise image is convolved with the
drizzled noise kernel given in \citeN{WBDDFFGGHKLLMPPAH96}, scaled to the
pixel size in the final drizzled image (by multiplying by 0.4), and
multiplied by a factor of 1.1 to account for the stochastic loss of
exposure time in each pixel due to cosmic-ray rejection. The
contribution from Poisson errors on counts from the objects is not
included, but is negligible for objects near the sky level.
The scaled noise image is then added to the noiseless object image
to produce the final simulated image. This prescription reproduces
the noise properties of the real HDF images remarkably well, as
can be seen by examination of the simulated images in Figures 
\ref{figimagehdf}-\ref{figimagelowq0}, and by detailed comparison
of image statistics on different scales. 
 
For quantitative comparisons, analysis of the images is carried
out using the FOCAS galaxy detection and 
photometry software (\citeNP{JT81,Valdes82}; revised 
by \citeNP{AS96}). The detection algorithm, as applied
to the HDF images, is discussed by \citeN{WBDDFFGGHKLLMPPAH96}.
Briefly, objects with S/N greater than $4\sigma$ within a
contiguous area of 25 pixels (each 0.04 square arcseconds in area) 
are considered
detections. Isophotal magnitudes are estimated by summing
the sky-subtracted counts within the detection isophote. 
For a point source on the WF, this minimum area encompasses
roughly 60\% of the total object flux. FOCAS total magnitudes
are computed from the total counts within an area that is
a factor of two larger.
The photometry of the simulated images and the HDF images
is identical, except for one minor difference. For the HDF
images, the individual pixels are weighted by the inverse
variance, to account for the small differences in exposure
time between adjacent pixels in the sub-sampled image. For
the simulations, the exposure time is assumed to be constant
for all pixels. This difference is likely to be unimportant,
as the variations in exposure time between pixels in the real
HDF images are typically less than 20\%.

In the case where there are multiple peaks within the initial
detection isophote, FOCAS computes photometric parameters
for both the ``parent'' and ``daughter'' objects. Because we
only want to count objects once, we have to decide, for 
each object, whether to keep the parent or the daughters. 
For both the true HDF images and the simulations, we have
adopted the separation and colour criteria of \citeN{WBDDFFGGHKLLMPPAH96},
keeping the parent if the daughters have $\Delta(V_{606}-I_{814}) < C$,
and separation $S_{ij} < F \times (r_i + r_j)$, with $C = 0.3$ mag
and $F = 5$, and $r_i$ and $r_j$ are galaxy radii determined from the FOCAS
isophotal areas as $r_i = \sqrt{A/\pi}$.
In the real HDF images, galaxy counts are only mildly
affected by the choice of whether to count parents or daughters,
growing by 20\% if all the subcomponents (daughters) rather
than the parents are counted
as individual objects \cite{WBDDFFGGHKLLMPPAH96}. 
The two main issues that may affect the comparison
of our models to the real images are (1) clustering on small
scales, and (2) substructure within galaxies. Galaxy positions in
our simulations are stastically independent. Intrinsic clustering
in the real universe is likely to lead to a reduction in
counts, as overlapping objects will in some cases be counted as
one object. The \citeN{VFdaC97} analysis of the angular correlation
function in the HDF suggests that clustering will lead to an excess
probability of about 10\% for galaxies to have a separation of 1 arcsec. 
Hence, we expect that clustering will lead to only a very slight 
undercounting of galaxies.
(Note that the higher probability found by \citeN{CGOR97} applies only to galaxies
with photometric redshifts $z > 2.4$, and their analysis used a different
algorithm for counting galaxies and merging subcomponents.)

Substructure within galaxies will work in the opposite direction. 
Galaxies in our models have smooth profiles, while real galaxies,
especially those being observed at rest-frame UV wavelengths, 
tend to have substructures. Even with the merging algorithm described
above, it is likely that some individual galaxies in the real HDF have been 
counted as multiple objects.  

These effects, while ultimately of great interest, have only 
a mild effect on the counts. The change in luminosities of
the individual components compensates for the change in the number
of components. Thus both the slope and normalization of the faint
counts stays approximately the same.
The effect on galaxy colors and size distributions is likely to be
larger, but we have not attempted to model it in this paper. 
The gross differences between the models and the
data, described below, are likely to be relatively insensitive to the 
details of the merging and splitting algorithms, although further
investigation is certainly warranted.

\section{Redshift Distributions} \label{sec3}
The $q_0=0.01$ model and the dwarf-dominated $q_0=0.5$ model, while
rather similar in their predicted redshift distributions to ground
based limits, differ dramatically in their predicted redshift
distributions at the depths probed by the HDF images. Figures
\ref{figzdw5} and \ref{figzlowq0} show the model redshift distributions
compared to the Canada-France redshift survey (CFRS; \citeNP{CFRS1}),
and the predicted redshift distributions in two fainter magnitude
slices. In the CFRS magnitude range ($17.5 < I_{\rm AB} < 22.5$), the
predicted distribution for both models provide a reasonable match to
the data for $z<2$ {\it given the incompleteness of the sample}. In the
$q_0 = 0.5$ model, 25\% of the galaxies (most right near the CFRS magnitude
limit) have $z > 2$, which is just compatible with the 81\%
completeness of the sample.  The high-redshift tail in this magnitude
range is much less pronounced for the low $q_0$ model because the
luminosity-distance is larger at high redshift. 


At fainter magnitudes, the dwarfs in the $q_0 = 0.5$ model dominate
the counts and the redshift distributions. While the vast majority
of the dwarfs form their stars at redshifts $0.5 < z < 1$ in the
model, the counts and redshifts are dominated by faded dwarf galaxies
at lower redshifts. This is root of the problems with matching 
the distributions of colour and radius described in the next section.
In the low $q_0$ model, the redshift distribution at HDF magnitudes
$24 < I_{814}< 27$ is dominated by high-redshift bulges and ellipticals. 
The redshift distribution is much more uniform at fainter magnitudes
$27 < I_{814}< 29$. The peak is missing at high redshift because 
most of the ellipticals and bulges are brighter than 
$I_{814} = 26$ during the epoch of rapid star formation. The ellipticals
and bulges that appear in the faint magnitude cut come from the 
faint tail of the adopted Gaussian luminosity function. For 
redshifts $z < 2$, faint sample is dominated by late-type 
low-luminosity galaxies, which in this model are forming stars
at roughly constant rates.

\section{Comparisons to HDF Observations} \label{sec2}

In Figures \ref{figimagedw5} and \ref{figimagelowq0}, we show
simulated F814W HDF images for the dwarf-dominated model and the low
$q_0$ model, respectively.  In Figure \ref{figimagehdf}, we show the
HDF image.  A visual comparison of the images reveals that the image for
the dwarf-dominated model appears to have more galaxies than in the HDF
image, this impression being primarily due to the large number of
extended, low surface brightness galaxies near the detection limit that
are not present in the HDF image.  The bulk of the galaxies in the HDF
image are more point-like.  In contrast, the faintest galaxies in the 
image for the
low $q_0$ model are similar in size to those in the HDF image.  However,
the simulated image does not appear to have as many galaxies. 

The model counts are compared to the observations in figures \ref{figcountsdw5}
and \ref{figcountslowq0},
where we show, for the model, the underlying counts based on true
total magnitudes (dashed line), and the measured counts based on FOCAS
isophotal magnitudes (solid line) and ``total'' magnitudes. 
Note the rather striking differences between the underlying counts and
the FOCAS measured counts for both models. These differences justify
our suspicion that selection and photometry biases are important at
faint magnitudes, and must be included for a fair assessment
of the predictions for {\it any} galaxy-evolution model. 

The $q_0 = 0.5$ dwarf-dominated model clearly overpredicts the counts 
at HDF magnitudes, even when selection biases are taken into account
(Fig \ref{figcountsdw5}).
The excess is the most striking in the $I_{814}$ band, although it
appears at faint magnitudes in all bands. The excess is largely due
to fading dwarf galaxies, rather than dwarf galaxies at the peak
of their starformation activity.
The model predictions
for the $q_0 = 0.01$ model are a better match, being largely successful
in the $I_{814}$ band, while underpredicting the counts at relatively
bright $B_{450}$ magnitudes. This latter discrepancy is related
to the large surface-density of high-redshift galaxies predicted by the model,
and shows up as well in the colour distributions discussed below.
Note that this discrepancy would not have been seen in earlier
PLE models, which did not include the attenuation of the intergalactic medium
\shortcite{YT88,GK95,PBZ96}.

The distribution of radii are compared in Figure \ref{figradii}.
The radii plotted are first-moment radii measured by FOCAS within
the detection isophotes.\footnote{
The FOCAS first-moment radius is defined as 
\begin{equation}
r_1 = \sum r I(x,y) / \sum I(x,y),
\end{equation}
where $I(x,y)$ is the intensity in each pixel within the 
detection isophote.}

The observed distribution peaks at 0.2 arcsec in the 
magnitude range $24 < V_{606} < 27$, and at about 0.15 arcsec
in the range $27 < V_{606} < 29$. These distributions of radii are
almost perfectly matched by both models at the brighter magnitudes, and
by the low $q_0$ model in the fainter magnitude bin. However, 
in spite of being dwarf-dominated, the
$q_0 = 0.5$ model predicts galaxies that are systematically {\it larger}
than observed at the faintest magnitudes. 

Figure \ref{figcolors} compares the $B_{450} - I_{814}$ colours for 
the models and data. While the peaks in the model colour distributions
roughly agree with the observations, neither model provides a very
good match to the overall distributions. The $q_0 = 0.5$ model has
a blue peak in the brighter magnitude bin that is not observed, while
it predicts a red tail similar to that seen in the real HDF images.
At fainter magnitudes ($27 < I_{814} < 29$) there is no blue peak, and hence
the model predicts
a narrower distribution than observed.
The $q_0 = 0.01$ model does not have enough
blue galaxies at the brighter magnitudes, and greatly overpredicts
the number of very red galaxies ($B_{450}-I_{814} > 2$). At fainter
magnitudes the distribution is a better match, but there is still
an excess of very red galaxies relative to the observations.

The differences reflect both the different redshift distributions
of the models, and the different proportions of galaxies undergoing
star-formation at moderate redshifts.
During the starburst phase, the dwarfs in the
$q_0 = 0.5$ model show up as nearly flat spectrum objects,
giving rise to the blue peak  in the
magnitude range $24 < I_{814} < 27$. There are few galaxies in the HDF
with colours this blue. At fainter magnitudes, the colour distribution
is dominated by faded, lower-redshift remnants of the starburst epoch. 
The blue tail is missing because there are no dwarf galaxies in the 
model with star-formation rates less than 1.8 $M_\odot$ per year.
This is due the cutoff of the dwarf-galaxy mass function at 
15 $\rm\, km s^{-1}$. Lower mass potentials are not deep
enough to retain $10^4$ K gas during the formation epoch. 
While the lower mass cutoff has a physical justification, the 
adopted star-formation efficiency of 10\%, and star-formation
timescale of $10^7$ years are not highly constrained. Thus it is
may be possible to achieve a better match to the colour distribution 
without revising the fundamental assumption that dwarf galaxies
are dominating the counts.

For the low $q_0$ model, the most serious discrepancy is the
prediction that there should be a large population of very red
galaxies, primarily in the brighter magnitude bin, but extending
also to the faintest magnitudes seen in the HDF. This red tail
is due to bulges and ellipticals forming stars at $z > 3.5$. 
It is an extremely robust prediction of the models that such galaxies
should have red $B_{450} - I_{814}$ colours, since the colours
are determined largely by the intrinsic Lyman limits in galaxies, and
by absorption due to intervening Lyman $\alpha$ clouds 
\cite{Madau95}. In the bright magnitude bin, the low $q_0$ model,
even ignoring the very red tail, is skewed to the red of the
observed distribution. It is apparently underabundant in star-forming
galaxies at moderate redshifts. At fainter magnitudes, the colour
distribution is a more reasonable match to the observations, with
a higher proportion of blue objects. This is in part explained
by the virtual disappearance of passively evolving ellipticals from
the sample at moderate redshifts, because the observations have gone
beyond the peak of the assumed Gaussian luminosity functions. The 
counts are thus dominated by star-forming galaxies. 

Figure \ref{fighdfdropouts}, reproduced from \citeN{MFDGSF96}, 
shows colour-colour diagrams for galaxies on the three WF chips
in the HDF. The dashed lines show regions designed to select
galaxies in the redshift range $2.3 < z < 3.5$ (in the 
$U_{300}-B_{450}$ vs. $B_{450}-I_{814}$ 
plane shown in Fig. \ref{fighdfdropouts}a) and $3.5 < z < 4.5$ 
(in the $B_{450}-V_{606}$ vs. $V_{606}-I_{814}$  plane
in Fig. \ref{fighdfdropouts}b). Galaxies within these regions
are very likely to be star-forming (relatively blue) galaxies
at high redshifts that have ``dropped out'' of the bluer band
due to Lyman limit and intergalactic Ly$\alpha$ absorption. The
number of dropouts seen and predicted by the models are listed
in Table \ref{tabdropouts}. It is interesting and important
to note that this prediction is {\it extremely sensitive}
to biases in the selection and photometry of galaxies within
FOCAS. Because the counts of high-redshift galaxies in both models 
are steep at faint magnitudes, the $\sim 0.5$ mag difference
between FOCAS isophotal magnitudes and the model galaxy total
magnitudes can introduce large differences in the predicted
number of dropout galaxies. These differences are 
largely taken into account by using FOCAS on the simulated images,
but our conclusions below must certainly be tempered by the 
lack of knowledge of the true surface-brightness profiles of
high-redshift galaxies.

The predictions of the dwarf-dominated $q_0 = 0.5$ model are shown
in Fig. \ref{figdw5dropouts}. The model underpredicts the number of
U-band dropouts and overpredicts the number of B-band dropouts.
The discrepancy is significantly worse for the low $q_0$ model (Fig.
\ref{figlowq0dropouts}), which
greatly underpredicts the number of U dropouts, and greatly overpredicts
the number of B dropouts. Furthermore, for both models, B dropouts
are seen at magnitudes considerably brighter than those observed.

The differences between the two models can be understood as follows.
The volume at high redshifts is considerably higher in the low $q_0$
model, leading to a higher surface density of objects projected on the 
sky. Thus, for the same local density and redshift of formation, a low
$q_0$ model predicts more sources near $z=5$ than a $q_0 = 0.5$ model.
However, there is also more time between redshifts of 3 and 5 in the
low $q_0$ model (1.6 Gyr) than in the $q_0 = 0.5$ model (0.7 Gyr). 
Therefore galaxies that form at $z = 5$ have more time to fade in 
the open model, and, if the star-formation timescale is sufficiently
short, can fade and redden sufficiently that they do not meet the 
U-dropout selection criteria. 
The specific predictions are very
sensitive to the assumed redshift of formation and duration of the 
star-forming epoch, points to which we will return in the next section.

There are significant discrepancies between the models and the 
observations in other parts of these diagrams as well. In particular,
during the burst phase, the dwarfs in the $q_0 = 0.5$ model populate
a region slightly blueward of flat spectrum (in the lower left corner
of Fig. \ref{figdw5dropouts}a). There are virtually no galaxies in this
region in the HDF data.

The heavy concentration of galaxies at 
($0 < U_{300}-B_{450} <  0.5; B_{450}-I_{814} \approx 0.5$) is not
present in either model. Galaxies 
in this portion of the diagram are likely to lie in the
redshift range $1 < z < 2$.

\section{Discussion}

As outlined in the introduction, our purpose in this paper is to
examine the constraints imposed on the bursting-dwarf model by the
Hubble Deep Field observations. By way of comparison, we perform the
same analysis on a standard low $q_0$ PLE model, to highlight the
differences between the predictions of the models, and the
discriminatory power of these deep images. Both types of 
models have been shown previously to
provide a reasonable fit to ground-based counts and redshift
distributions. The preceding sections have presented a detailed comparison of
these models to the HDF data. The comparisons
have highlighted (a) the importance of including selection biases
into the comparisons of models to the HDF data, (b) the 
large differences in the predicted counts and colour distributions
for these two rather different models, and (c) the failure of either
model to reproduce the observations.

The most important discrepancy for the low $q_0$ PLE model is the
prediction of a substantial number of $z > 3.5$ galaxies at relatively
bright magnitudes. Previous PLE models \cite{GK95,PBZ96}
have shown reasonable agreement with ground-based data in the magnitude
range $24 < B < 26$, where we see the discrepancy, 
but this may have been fortuitous, as these models did
not include the effects of intergalactic absorption, which is clearly
important at faint magnitudes in the $U$ and $B$ bands. The number and
brightness distribution of high-redshift galaxies is sensitive to the
redshift of formation $z_f$, the star-formation timescale, the stellar
initial mass function, and the amount and distribution of dust in young
galaxies. Thus it is not possible to rule out PLE models simply on the
basis of the overprediction of $z > 3.5$ galaxies. However, simply
hiding these galaxies is not sufficient to reconcile the model to the
observations because then the PLE model will substantially underpredict
the counts. The most straightforward solution, of course, is to 
posit different redshifts of formation for different types of galaxies, 
or to incorporate merging into the model. Such possibilities 
are certainly viable, but are no longer in the spirit of PLE models.

The disagreement between the $q_0 = 0.5$ dwarf-dominated model and the
HDF observations can be largely attributed to the overproduction of
faded remnants. The model counts near the HDF limits are 
dominated by low-redshift non-starforming dwarfs. These remnants are
both redder and larger than the typical galaxies seen in the HDF. There
are two plausible modifications to the model which might reconcile it
to the observations without contradicting the underlying assumptions of
$q_0 = 0.5$, a CDM mass-spectrum for the dwarfs, and a redshift of
formation governed by the ionisation history of the universe.  The
first is that the formation epoch of the dwarfs could extend to higher
redshifts than we have assumed (possibly up to $z = 1.5$), and could
depend on galaxy mass. This would allow the boojums to be brighter and
leave them more time to fade.  The second is that the initial mass
function in the boojums could be skewed toward high mass stars, or
truncated at some fairly high lower mass limit. The skewed IMF would
cause the dwarfs to fade faster, possibly curing the problem of
remnants at low redshifts.

In a broader sense, the HDF observations clearly provide new and
important constraints on galaxy-evolution models. At the faint
magnitudes probed by these observations, the problems with the simple
models discussed in this paper are particularly acute. It is difficult
to tell at present whether the problems lie in the details (e.g. the
IMF, dust content, metallicity distributions) or in the fundamental
assumptions (e.g. that merging is unimportant, or that giant galaxies
all began forming at roughly the same time, with different star-forming
rates). It is clear that further modeling of the growing database of
galaxy properties in this small patch of sky, with careful attention
to selection effects, has the potential to discriminate between widely
different world models that heretofore seemed equally plausible.

This paper is based on observations made with the Hubble Space
Telescope; support for this work was provided in part by NASA
through Hubble Archival Research grant \#AR-06337.20-94A
awarded by Space Telescope Science Institute, which is operated
by the Association of Universities for Research in Astronomy, Inc.,
under NASA contract NAS5-26555. A.B. is grateful for support through
the Bergen Career Award from the Dudley Observatory. We would
like to acknowledge Bob Williams and the STScI HDF team for their
efforts in planning and carrying out the HDF observations. We 
acknowledge in particular Mark Dickinson, Andy Fruchter, Mauro
Giavalisco, and Piero Madau for many fruitful discussions and
collaborations on the HDF and galaxy evolution. 

\begin{table}
\begin{center}
\caption{Giant Galaxy Parameters
\label{tabgalpars}
}
\begin{tabular}{lcccccccc}
\multicolumn{1}{c}{} &
\multicolumn{3}{c}{Luminosity Function} &
\multicolumn{2}{c}{Bulge/Total Ratio} \\
Type & $N_0 (\times 10^{-3} \rm Mpc^{-3}$) & $<M_{\rm BJ}>$ & $\sigma$ & mean & $\sigma$ & $z_f$ & $\tau_{\rm Bulge}$ & $\tau_{\rm Disk}$\\
\hline
E & 0.37 & -18.70 & 1.7 & 1.0 & 0.0 & 5 & 0.2 & ---\\
S0 & 1.15 & -18.70 & 1.7 & 0.4 & 0.25 & 5 & 0.2 & 1 \\
Sab & 2.00 & -19.96 & 1.1 & 0.3 & 0.27 & 5 & 0.2 & 30\\
Sbc & 4.00 & -18.48 & 1.3 & 0.15 & 0.27 & 5 & 0.2 & 30\\
Sdm & 8.00 & -16.00 & 1.3 & 0.0 & 0.0 & 5 & --- & 30\\
\\
\multicolumn{8}{c}{Additional late-type dwarfs for low $q_0$ model} \\
Type & $N_0 (\times 10^{-3} \rm Mpc^{-3}$) & $M_{BJ}^*$ & $\alpha$ & mean & $\sigma$ & $z_f$ & $\tau_{\rm Bulge}$ & $\tau_{\rm Disk}$\\
\hline
Irregular & 200 & -16.00 & -1.3 & 0.0 & 0.0 & 5 & --- & 30\\
\\
\multicolumn{8}{c}{Bursting dwarfs for $q_0 = 0.5$ model} \\
Type & $N_0 (\times 10^{-3} \rm Mpc^{-3}$) & $M_0$ & $\alpha$ & mean & $\sigma$ & $z_{\rm max}$ & $\tau_{\rm Bulge}$ &SF duration\\
\hline
Boojum & 4000 & $10^9 M_\odot$ & -2 & 0.0 & 0.0 & 1 & --- & 0.01\\
\hline
\end{tabular}
\end{center}
\end{table}

\begin{table}
\begin{center}
\caption{Numbers of $U_{300}$ and $B_{450}$ band Dropouts
\label{tabdropouts}
}
\begin{tabular}{lcc}
Data set & U dropouts & B dropouts \\
\hline
HDF & 58 & 14 \\
$q_0 = 0.5$ & 30 & 41 \\
$q_0 = 0.01$ & 2 & 172 \\
\hline
\end{tabular}
\end{center}
\end{table}

\clearpage
\bibliography{mnrasmnemonic,bib} 

\begin{thebibliography}{}

\bibitem[\protect\citeauthoryear{{Adelberger} \& {Steidel}}{{Adelberger} \&
  {Steidel}}{1996}]{AS96}
{Adelberger} K.,  {Steidel} C.~C., 1996, private communication

\bibitem[\protect\citeauthoryear{{Babul} \& {Ferguson}}{{Babul} \&
  {Ferguson}}{1996}]{BF96}
{Babul} A.,  {Ferguson} H.~C., 1996, ApJ, 458, 100

\bibitem[\protect\citeauthoryear{{Babul} \& {Rees}}{{Babul} \&
  {Rees}}{1992}]{BR92}
{Babul} A.,  {Rees} M., 1992, MNRAS, 255, 346

\bibitem[\protect\citeauthoryear{{Bertelli} et~al.}{{Bertelli}
  et~al.}{1994}]{BBCFN94}
{Bertelli} G., {Bressan} A., {Chiosi} C., {Fagotto} F.,  {Nasi} E., 1994,
  A\&AS, 106, 275

\bibitem[\protect\citeauthoryear{{Binggeli}, {Sandage}, \&
  {Tammann}}{{Binggeli} et~al.}{1988}]{BST88annrev}
{Binggeli} B., {Sandage} A.,  {Tammann} G.~A., 1988, ARA\&A, 26, 509

\bibitem[\protect\citeauthoryear{{Biretta}}{{Biretta}}{1995}]{WFPC2handbook}
{Biretta} J., 1995, WFPC-2 Instrument Handbook.
\newblock STScI, Baltimore

\bibitem[\protect\citeauthoryear{{Cole} et~al.}{{Cole} et~al.}{1994}]{CAFNZ94}
{Cole} S., {Aragon-Salamanca} A., {Frenk} C.~S., {Navarro} J.~F.,  {Zepf}
  S.~E., 1994, MNRAS, 271, 781

\bibitem[\protect\citeauthoryear{{Colley} et~al.}{{Colley}
  et~al.}{1997}]{CGOR97}
{Colley} W., {Gnedin} O., {Ostriker} J.~P.,  {Rhoads} J.~E., 1997, ApJ, 488,
  579

\bibitem[\protect\citeauthoryear{{Cowie} et~al.}{{Cowie} et~al.}{1996}]{CSHC96}
{Cowie} L.~L., {Songaila} A., {Hu} E.~M.,  {Cohen} J.~G., 1996, AJ, 112, 839

\bibitem[\protect\citeauthoryear{{{De Propis}} et~al.}{{{De Propis}}
  et~al.}{1995}]{DePPHM95}
{{De Propis}} R., {Pritchett} C.~J., {Harris} W.~E.,  {McClure} R.~D., 1995,
  ApJ, 450, 534

\bibitem[\protect\citeauthoryear{{Ellis} et~al.}{{Ellis}
  et~al.}{1996}]{ECBHG96}
{Ellis} R.~S., {Colless} M., {Broadhurst} T., {Heyl} J.,  {Glazebrook} K.,
  1996, MNRAS, 280, 235

\bibitem[\protect\citeauthoryear{{Gibson}}{{Gibson}}{1996}]{Gibson96}
{Gibson} B., 1996, ApJ, 468, 167

\bibitem[\protect\citeauthoryear{{Glazebrook} et~al.}{{Glazebrook}
  et~al.}{1995}]{GECBAT95}
{Glazebrook} K., {Ellis} R., {Colless} M., {Broadhurst} T., {Allington-Smith}
  J.,  {Tanvir} N., 1995, MNRAS, 273, 157

\bibitem[\protect\citeauthoryear{{Gronwall} \& {Koo}}{{Gronwall} \&
  {Koo}}{1995}]{GK95}
{Gronwall} C.,  {Koo} D.~C., 1995, ApJ, 440, L1

\bibitem[\protect\citeauthoryear{{Jarvis} \& {Tyson}}{{Jarvis} \&
  {Tyson}}{1981}]{JT81}
{Jarvis} J.~F.,  {Tyson} J.~A., 1981, AJ, 86, 476

\bibitem[\protect\citeauthoryear{{Kauffmann}, {Guiderdoni}, \&
  {White}}{{Kauffmann} et~al.}{1994}]{KGW94}
{Kauffmann} G., {Guiderdoni} B.,  {White} S.~D.~M., 1994, MNRAS, 267, 981

\bibitem[\protect\citeauthoryear{{Lilly} et~al.}{{Lilly} et~al.}{1995}]{CFRS1}
{Lilly} S.~J., {{Le Fevre}} O., {Crampton} D., {Hammer} F.,  {Tresse} L., 1995,
  ApJ, 455, 50

\bibitem[\protect\citeauthoryear{{Lin} et~al.}{{Lin} et~al.}{1996}]{LKSLOTS96}
{Lin} H., {Kirshner} R.~P., {Schectman} S.~A., {Landy} S.~D., {Oemler} A.,
  {Tucker} D.~L.,  {Schechter} P.~L., 1996, ApJ, 464, 60

\bibitem[\protect\citeauthoryear{{Loveday} et~al.}{{Loveday}
  et~al.}{1992}]{LPEM92}
{Loveday} J., {Peterson} B.~A., {Efstathiou} G.,  {Maddox} S., 1992, ApJ, 390,
  338

\bibitem[\protect\citeauthoryear{{Madau}}{{Madau}}{1995}]{Madau95}
{Madau} P., 1995, ApJ, 441, 18

\bibitem[\protect\citeauthoryear{{Madau} et~al.}{{Madau}
  et~al.}{1996}]{MFDGSF96}
{Madau} P., {Ferguson} H.~C., {Dickinson} M., {Giavalisco} M., {Steidel} C.~C.,
   {Fruchter} A.~S., 1996, MNRAS, 283, 1388

\bibitem[\protect\citeauthoryear{{Marzke}, {Huchra}, \& {Geller}}{{Marzke}
  et~al.}{1994}]{MHG94}
{Marzke} R.~O., {Huchra} J.~P.,  {Geller} M.~J., 1994, ApJ, 428, 43

\bibitem[\protect\citeauthoryear{{Phillipps} \& {Driver}}{{Phillipps} \&
  {Driver}}{1995}]{PD95}
{Phillipps} S.,  {Driver} S., 1995, MNRAS, 274, 832

\bibitem[\protect\citeauthoryear{{Pozzetti}, {Bruzual}, \&
  {Zamorani}}{{Pozzetti} et~al.}{1996}]{PBZ96}
{Pozzetti} L., {Bruzual} G.,  {Zamorani} G., 1996, MNRAS, 281, 953

\bibitem[\protect\citeauthoryear{{Press} \& {Schechter}}{{Press} \&
  {Schechter}}{1974}]{PS74}
{Press} W.~H.,  {Schechter} P.~L., 1974, ApJ, 187, 425

\bibitem[\protect\citeauthoryear{{Quinn}, {Katz}, \& {Efstathiou}}{{Quinn}
  et~al.}{1996}]{QKE96}
{Quinn} T., {Katz} N.,  {Efstathiou} G., 1996, MNRAS, 278, L41

\bibitem[\protect\citeauthoryear{{Rocca-Volmerange} \&
  {Guiderdoni}}{{Rocca-Volmerange} \& {Guiderdoni}}{1988}]{RG88}
{Rocca-Volmerange} B.,  {Guiderdoni} B., 1988, A\&AS, 75, 93

\bibitem[\protect\citeauthoryear{{Schechter}}{{Schechter}}{1976}]{Schechter76}
{Schechter} P., 1976, ApJ, 203, 297

\bibitem[\protect\citeauthoryear{{Simien} \& {{de Vaucouleurs}}}{{Simien} \&
  {{de Vaucouleurs}}}{1986}]{SdV86}
{Simien} F.,  {{de Vaucouleurs}} G., 1986, ApJ, 302, 564

\bibitem[\protect\citeauthoryear{{Valdes}}{{Valdes}}{1982}]{Valdes82}
{Valdes} F., 1982, Faint Object Classification and Analysis System.
\newblock KPNO Internal Publication

\bibitem[\protect\citeauthoryear{{Villumsen}, {Freudling}, \& {{da
  Costa}}}{{Villumsen} et~al.}{1997}]{VFdaC97}
{Villumsen} J., {Freudling} W.,  {{da Costa}} L.~N., 1997, ApJ, 481, 578

\bibitem[\protect\citeauthoryear{{Williams} et~al.}{{Williams}
  et~al.}{1996}]{WBDDFFGGHKLLMPPAH96}
{Williams} R.~E. et~al., 1996, AJ, 112, 1335

\bibitem[\protect\citeauthoryear{{Yoshii} \& {Peterson}}{{Yoshii} \&
  {Peterson}}{1994}]{YP94}
{Yoshii} Y.,  {Peterson} B., 1994, ApJ, 436, 551

\bibitem[\protect\citeauthoryear{{Yoshii} \& {Takahara}}{{Yoshii} \&
  {Takahara}}{1988}]{YT88}
{Yoshii} Y.,  {Takahara} F., 1988, ApJ, 326, 1

\end{thebibliography}
\bibliographystyle{mnras}

\clearpage

\begin{figure}

\caption{\label{figimagehdf}
Hubble Deep Field. The image shows an $80" \times 80"$ portion of the field, in the
F814W band, from the version 2 drizzled data (Williams et al. 1996).
}

\caption{\label{figimagedw5}
Simulated Hubble Deep Field for the $q_0 = 0.5$ dwarf-dominated model. 
The image shows an $80" \times 80"$ portion of the field in the F814W band. 
}

\caption{\label{figimagelowq0}
Simulated Hubble Deep Field in the low-$q_0$ model. 
The image shows an $80" \times 80"$ portion of the field in the F814W band. 
}

\caption{\label{figzdw5}
Redshift distribution for the $q_0 = 0.5$ dwarf-dominated model. The top
panel shows the measured redshift distribution from the CFRS survey
(Lilly et al. 1995) as solid dots, the predicted distribution from the
dwarf-dominated model as a histogram, and the prediction of a standard
no-evolution model as a dotted curve. For the 
dwarf-dominated model the predicted distribution includes observational
selection as described by Babul \& Ferguson (1996). The model distributions
are normalized to have the same total number of galaxies as the survey.
The middle panel shows the predicted redshift distribution from the
dwarf-dominated model in the magnitude range $24 < I_{814} < 27$. The
bottom panel shows the same for $27 < I_{814} < 29$.
}

\caption{\label{figzlowq0}
Same as Figure 4, for the low-$q_0$ PLE model.
}

\caption{\label{figcountsdw5}
Galaxy counts for the $q_0 = 0.5$ model
as a function of AB magnitude in the F450W and F814W bands,
together with a compilation of existing ground-based data.
FOCAS isophotal and total magnitudes are shown as solid dots
and X's, respectively. Model predictions based on total magnitudes
(i.e. {\it without} accounting for observational selection or photometric
biases) are shown as dashed lines. Model predictions from the 
FOCAS-generated catalogs of simulated galaxies are shown as solid lines.
No colour corrections have been applied to the ground-based data. For the
typical colours of galaxies in the HDF, the colour corrections are less
than 0.1 mag. 
}

\caption{\label{figcountslowq0}
Same as Fig. 6, for the low $q_0$ model.
}

\end{figure}
\begin{figure}

\caption{\label{figradii}
Distribution of radii for both models, compared to the data. The
radii for both the models and the data are the 
first-moment radii measured by FOCAS. The HDF measurements
are shown as histograms with Poisson error bars. The model predictions
are the light solid curves. 
}

\caption{\label{figcolors}
Distribution of $B_{450}-I_{814}$ colours for both models, compared to the data. 
The colours for both the models and the data are from FOCAS measurements
of the images. The HDF measurements
are shown as histograms with Poisson error bars. The model predictions
are the light solid curves. For both the models and the data there are
galaxies that are detected in the F814W image but not in the F450W image.
For these galaxies, the colour is assigned from the $1\sigma$ lower limit
to the $B_{450}$ magnitude. The hashed regions indicate portion of 
the diagrams populated by these lower limits. Hashes slant in opposite directions
for the models and the data.
}

\caption{\label{fighdfdropouts}
(a) $U_{300} - B_{450}$ vs. $B_{450} - I_{814}$ colour-colour plot of galaxies in
the HDF with $B_{450} < 26.8$. Objects undetected in F300W (with signal-to-noise $< 1$)
are plotted as triangles at the $1\sigma$ lower limits to their $U_{300} - B_{450}$
colours. Symbol size scales with the $I_{814}$ magnitude of each object.
The dashed lines outline the selection region within which we identify
candidate $2 < z < 3.5$ objects. Details of the selection criteria are given
by Madau et al. (1996). 
(b) $B_{450} - V_{606}$ vs. $V_{606} - I_{814}$ colour-colour plot of galaxies
in the HDF with $V_{606} < 28.0$. Meanings of the symbols are the same as
in the previous plot. The dashed lines bound the region that isolates galaxies
with $3.5 < z < 4.5$.
}

\caption{\label{figdw5dropouts}
Same as Figure 10, for the $q_0 = 0.5$ dwarf-dominated model.
Galaxy magnitudes are from FOCAS measurements, as for the data.
}

\caption{\label{figlowq0dropouts}
Same as Figure 11, for the $q_0 = 0.01$ PLE model.
Galaxy magnitudes are from FOCAS measurements, as for the data.
}
\end{figure}
\clearpage
Figs 1-3 are in separate files.
\clearpage
Figure 4.
\plotone{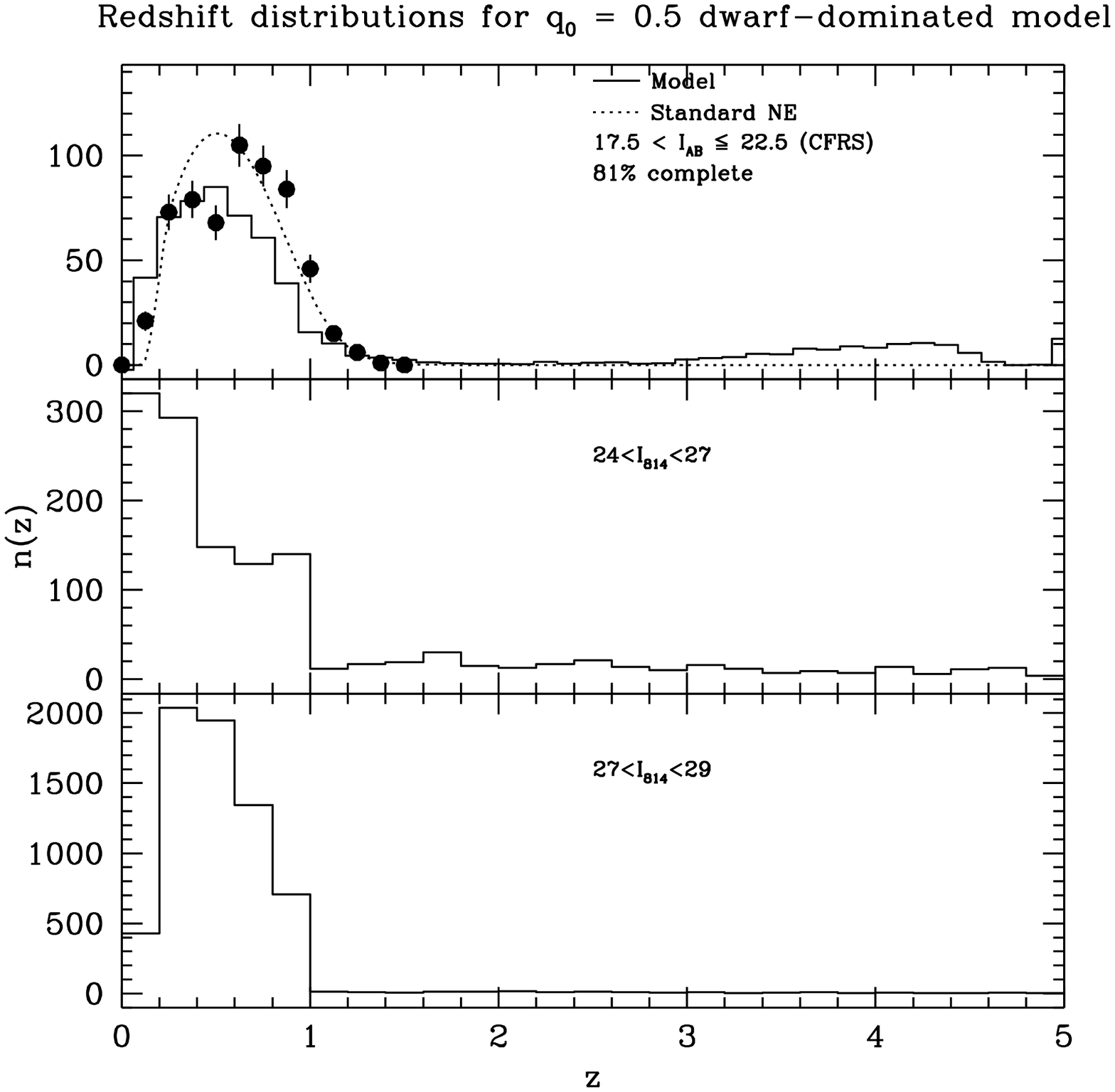}
\clearpage
Figure 5.
\plotone{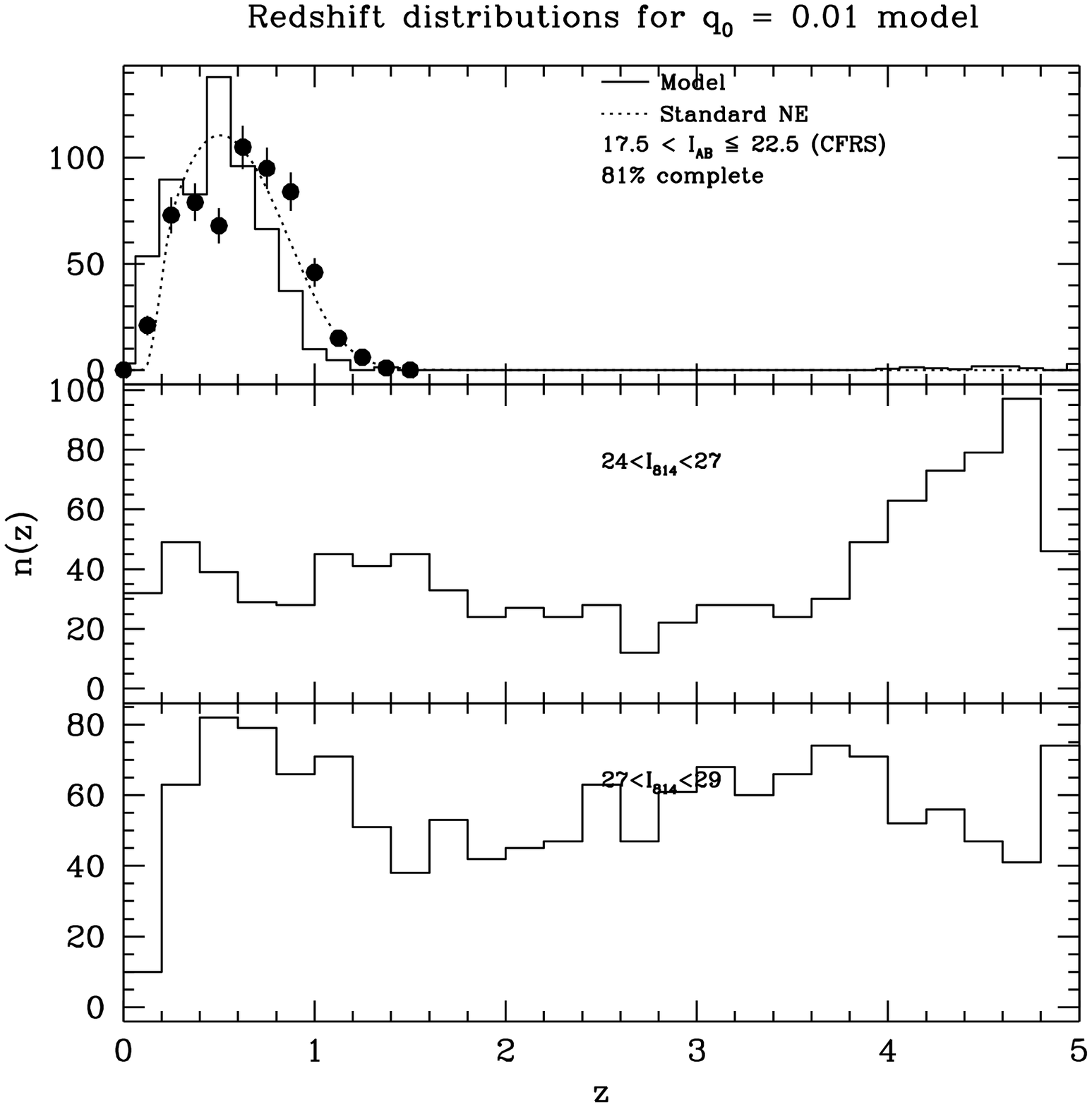}
\clearpage
Figure 6.
\plotone{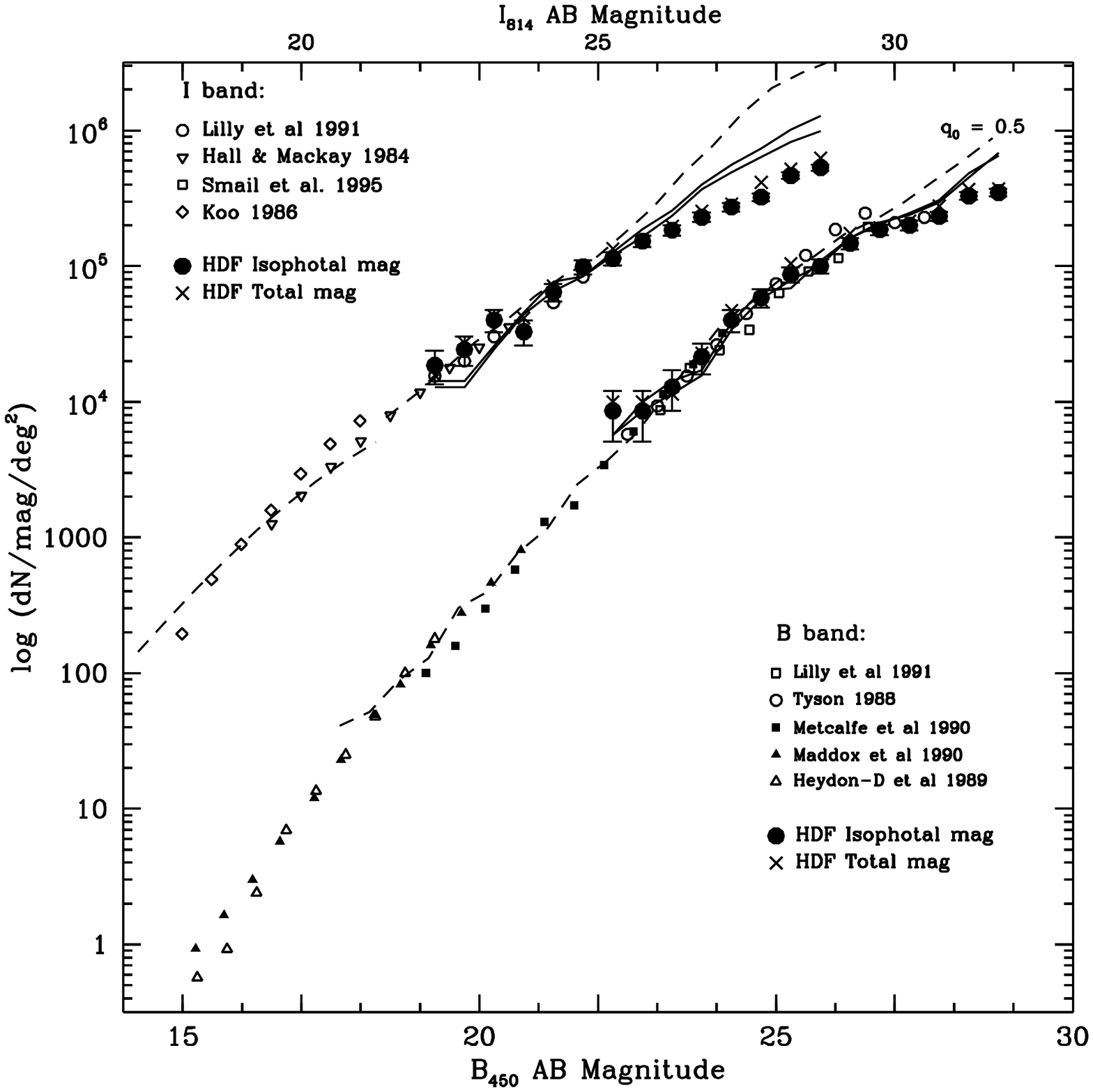}
\clearpage
Figure 7.
\plotone{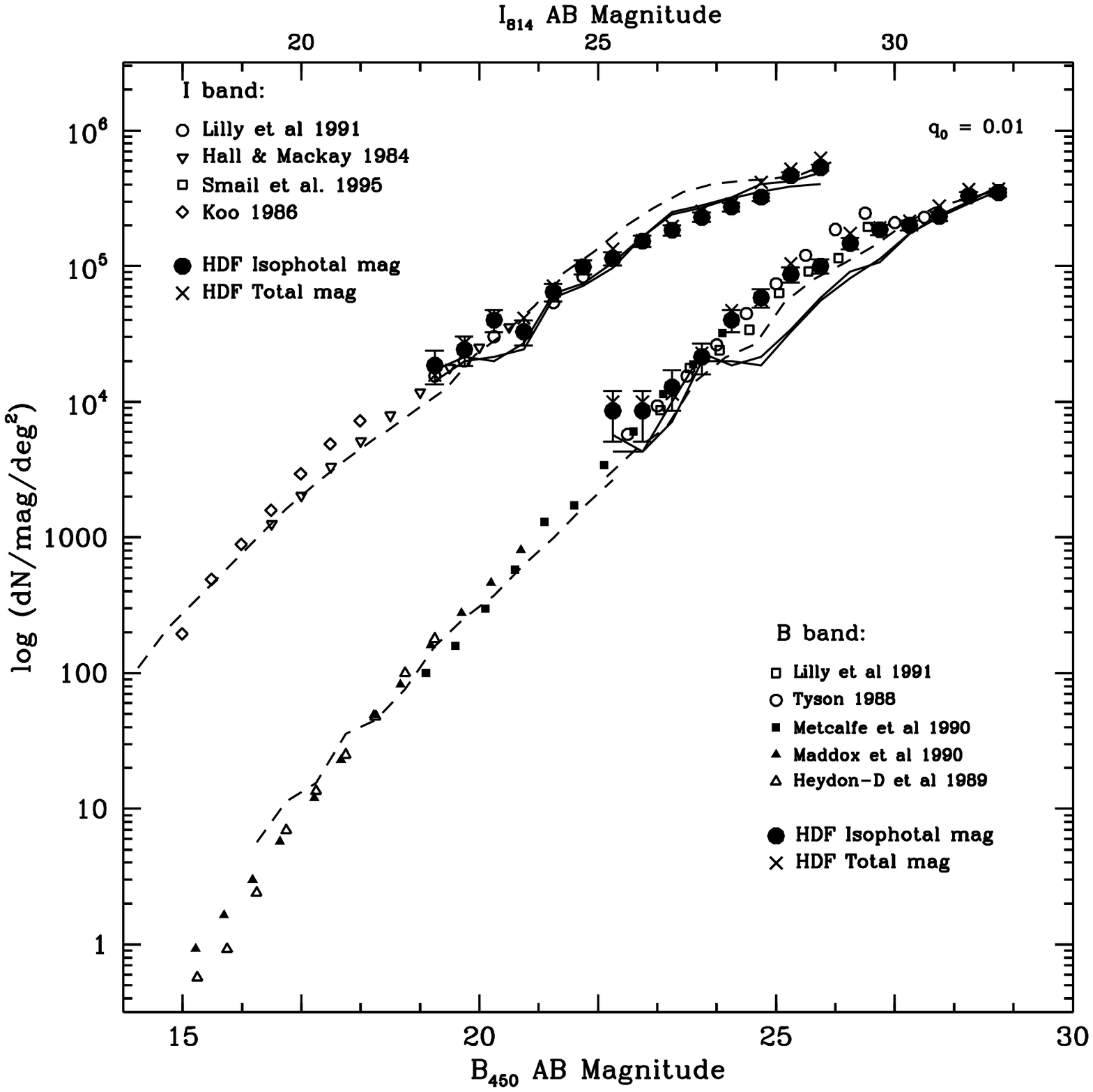}
\clearpage
Figure 8.
\plotone{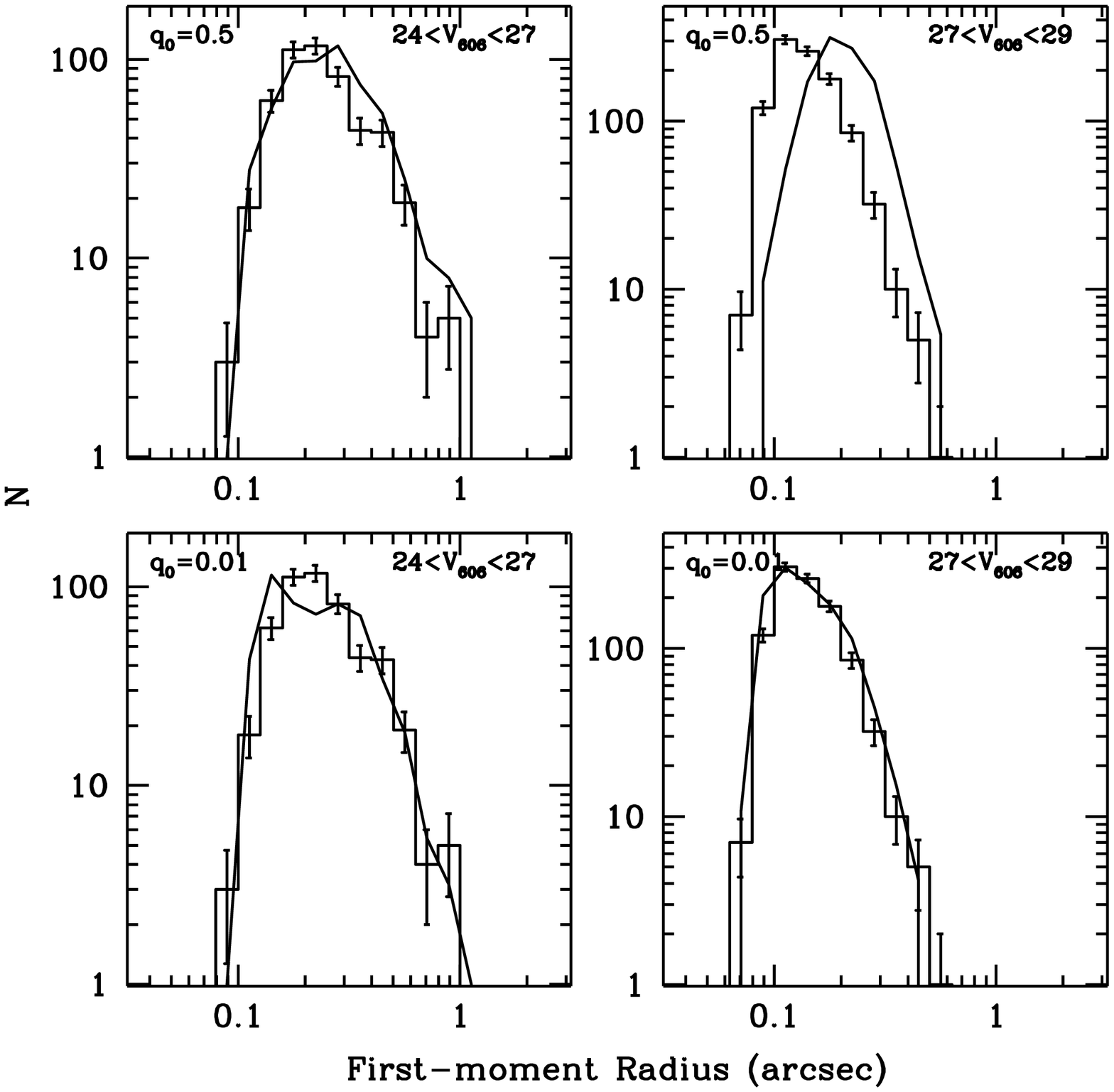}
\clearpage
Figure 9.
\plotone{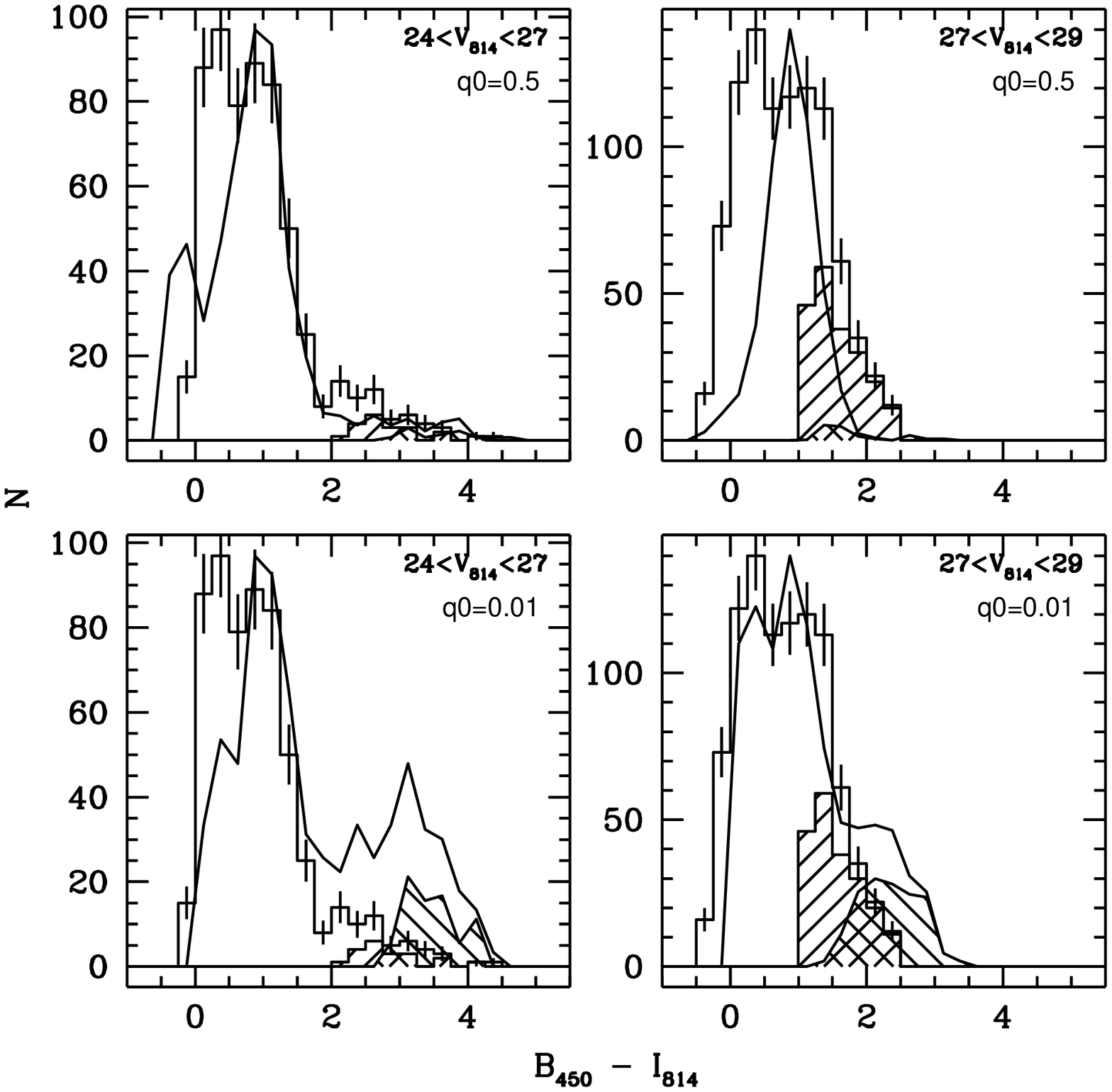}
\clearpage
Figure 10a.
\plotone{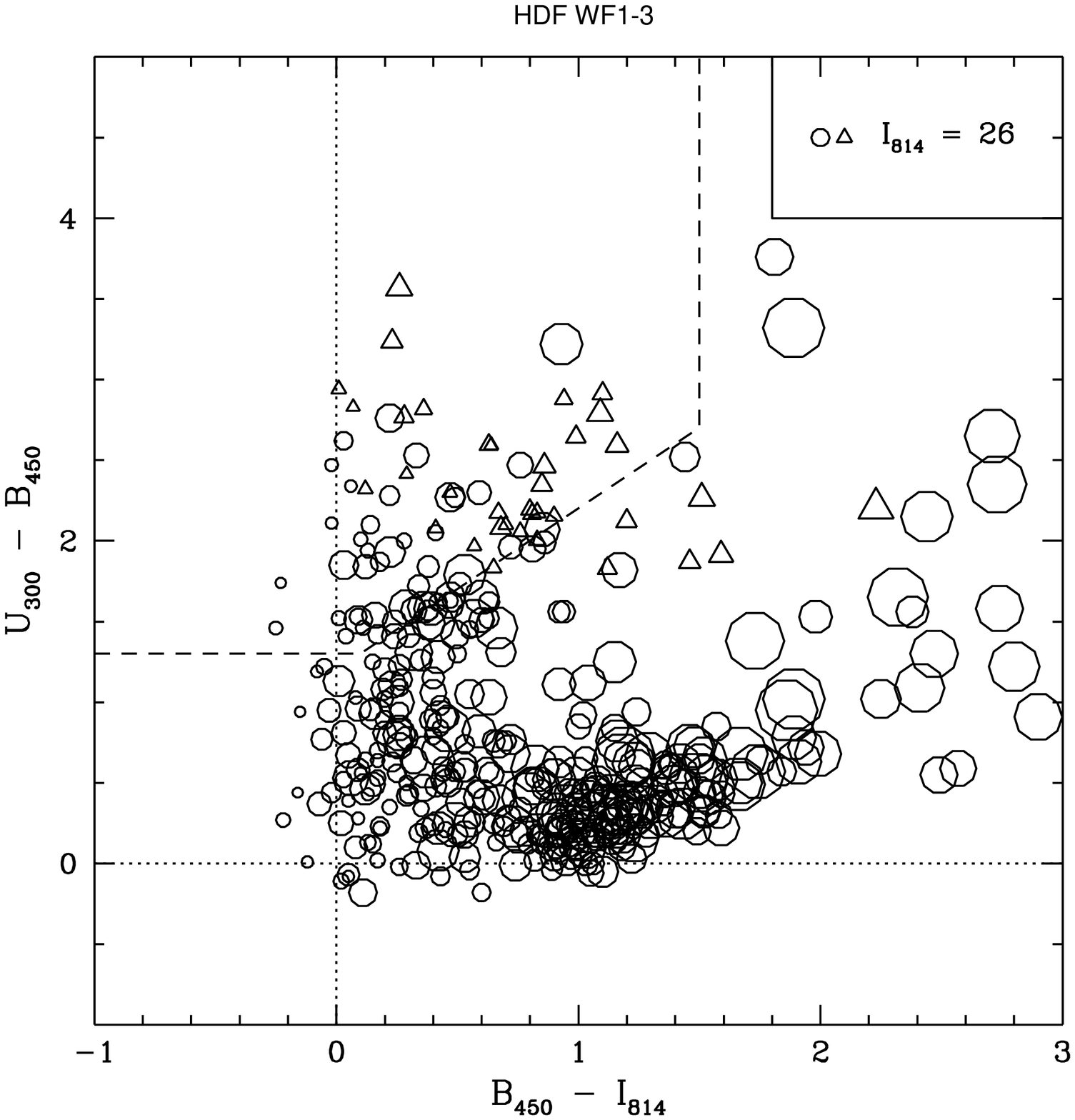}
\clearpage
Figure 10b.
\plotone{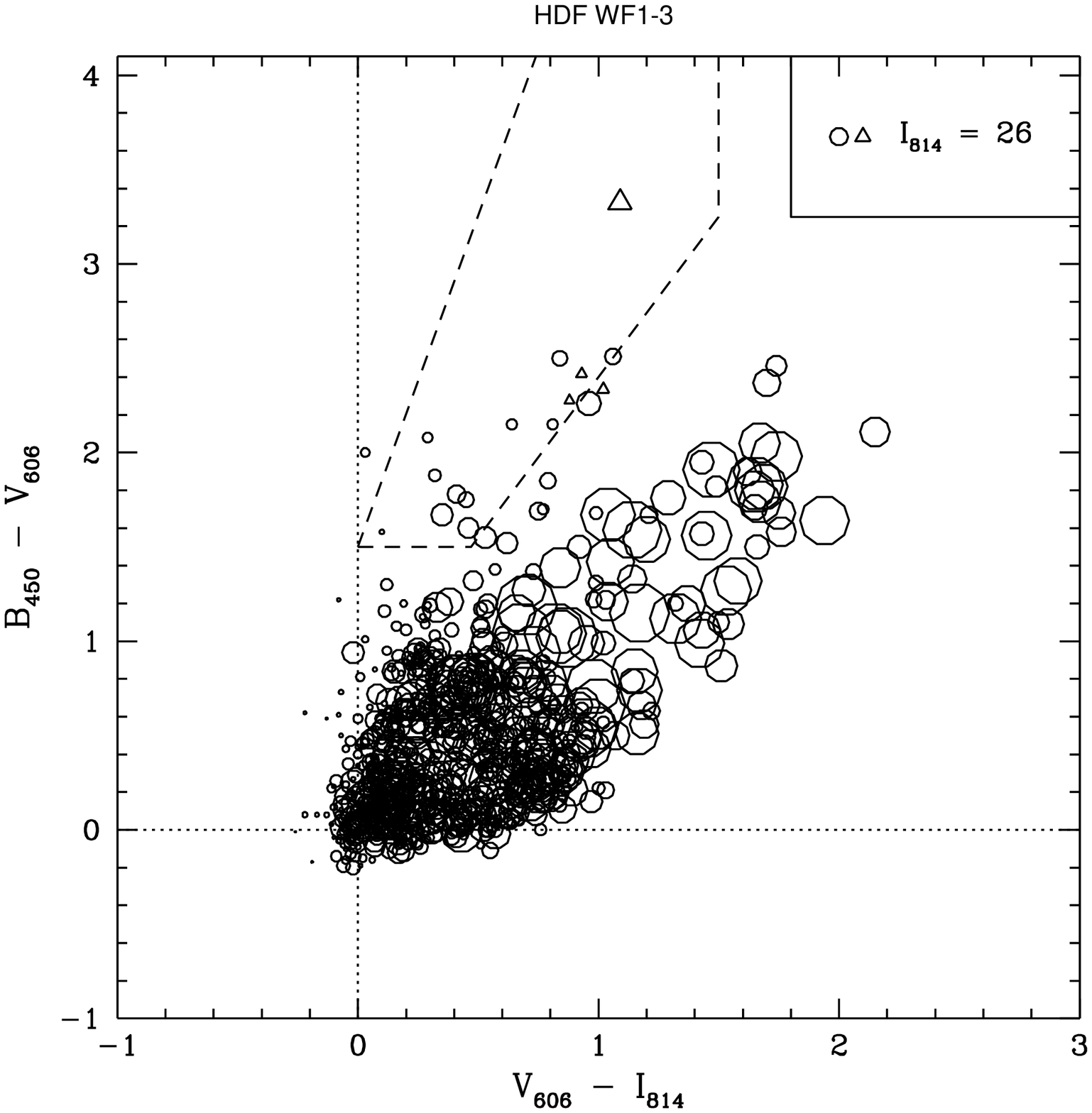}
\clearpage
Figure 11a.
\plotone{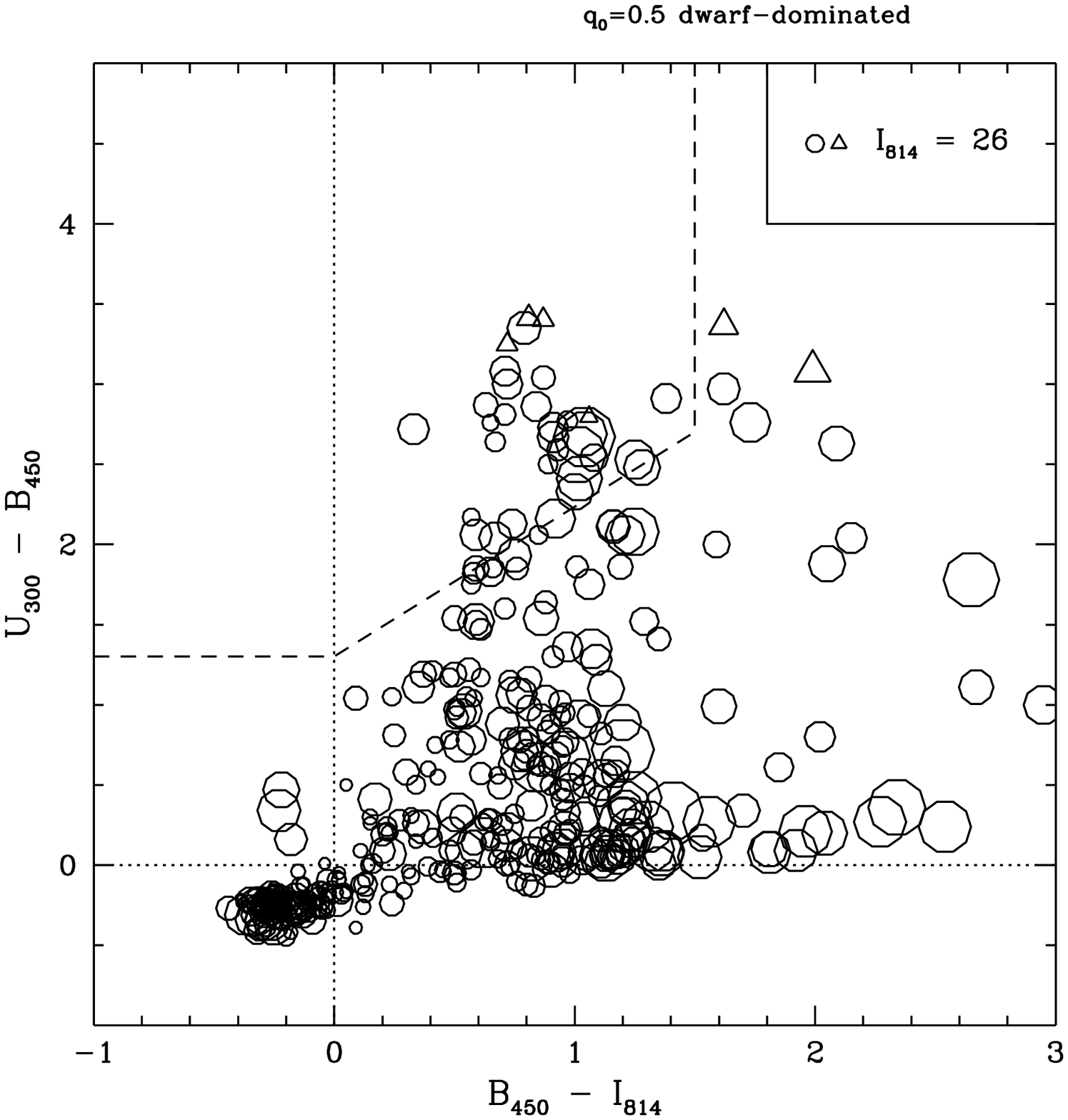}
\clearpage
Figure 11b.
\plotone{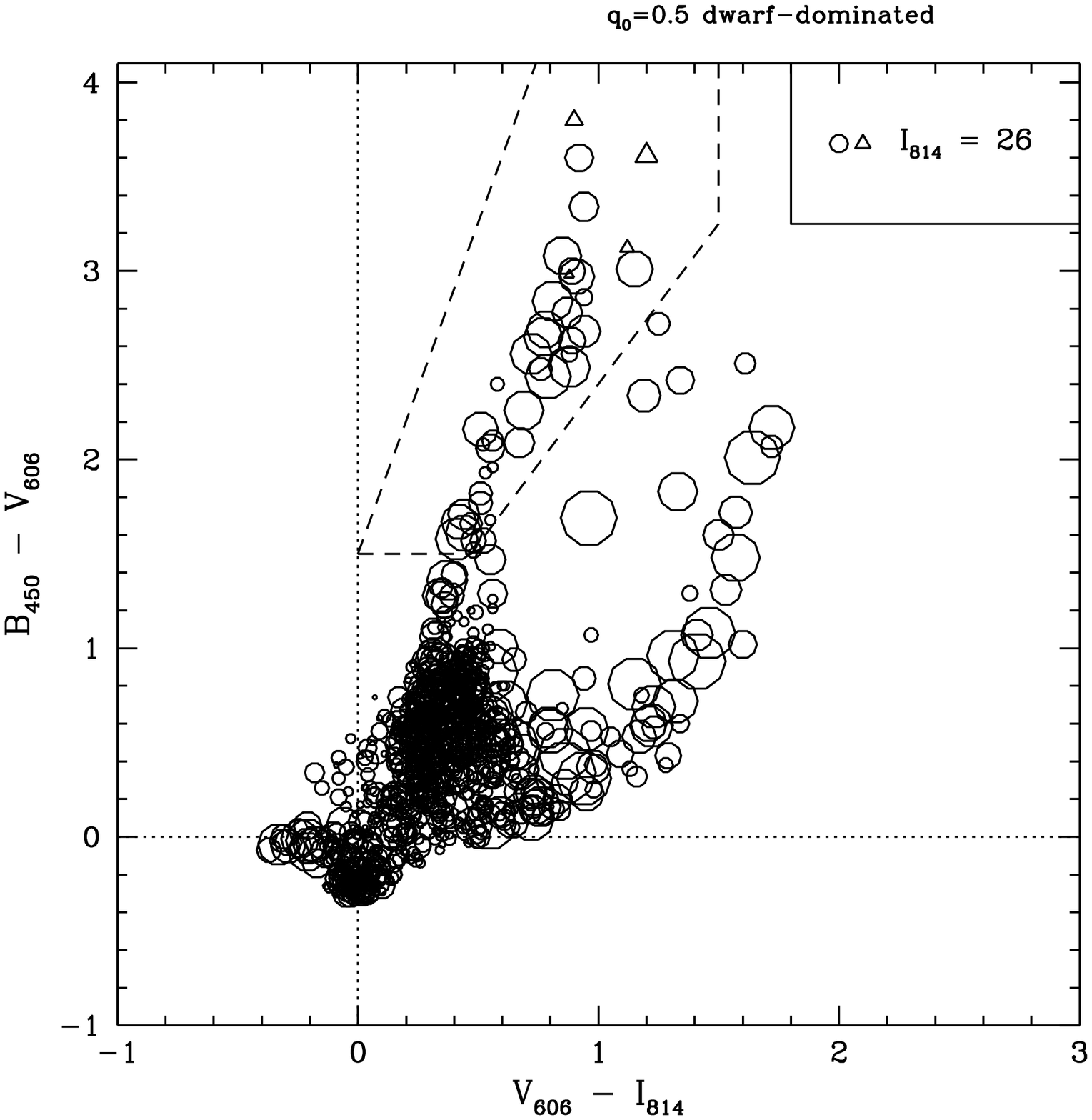}
\clearpage
Figure 12a.
\plotone{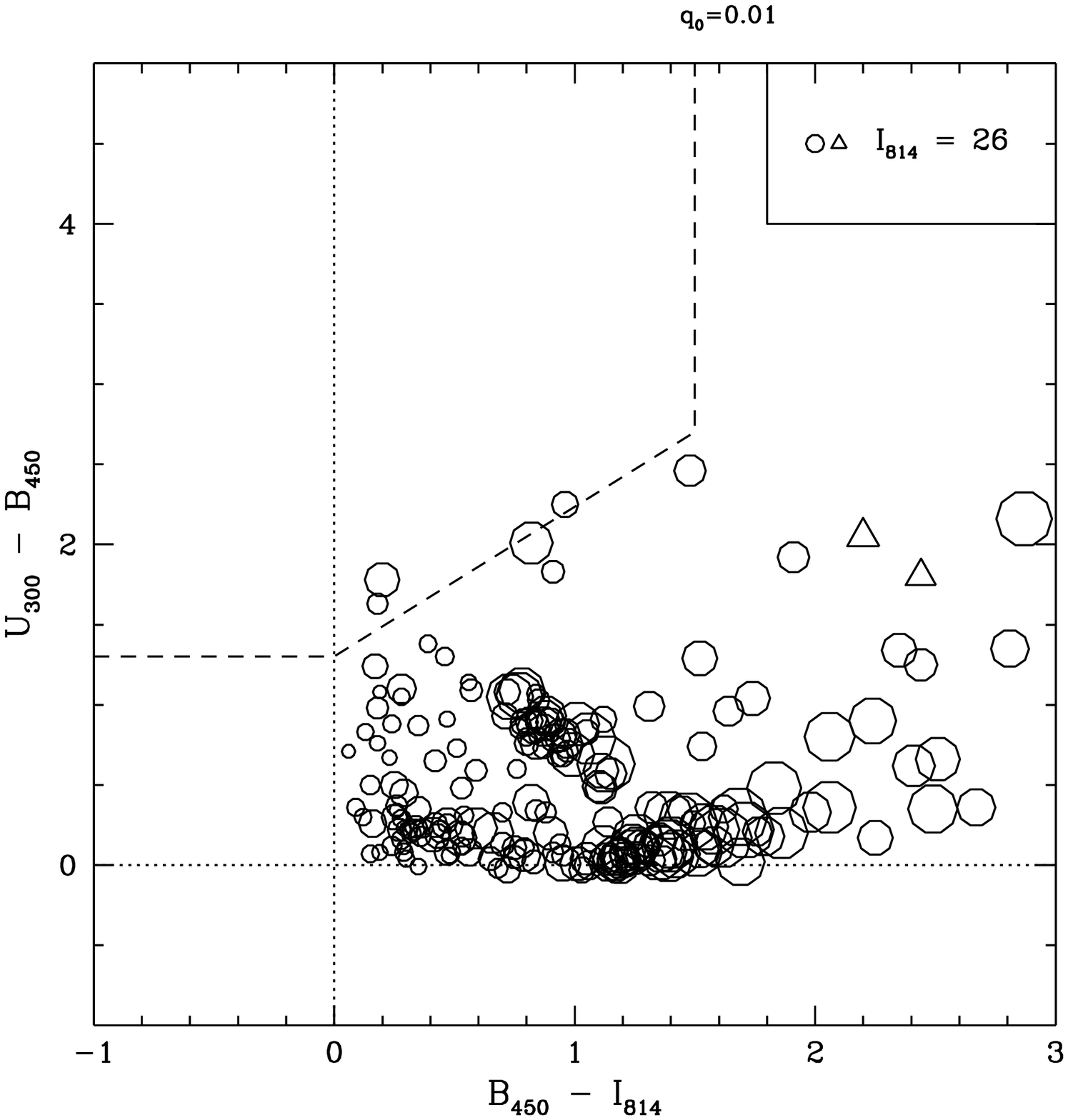}
\clearpage
Figure 12b.
\plotone{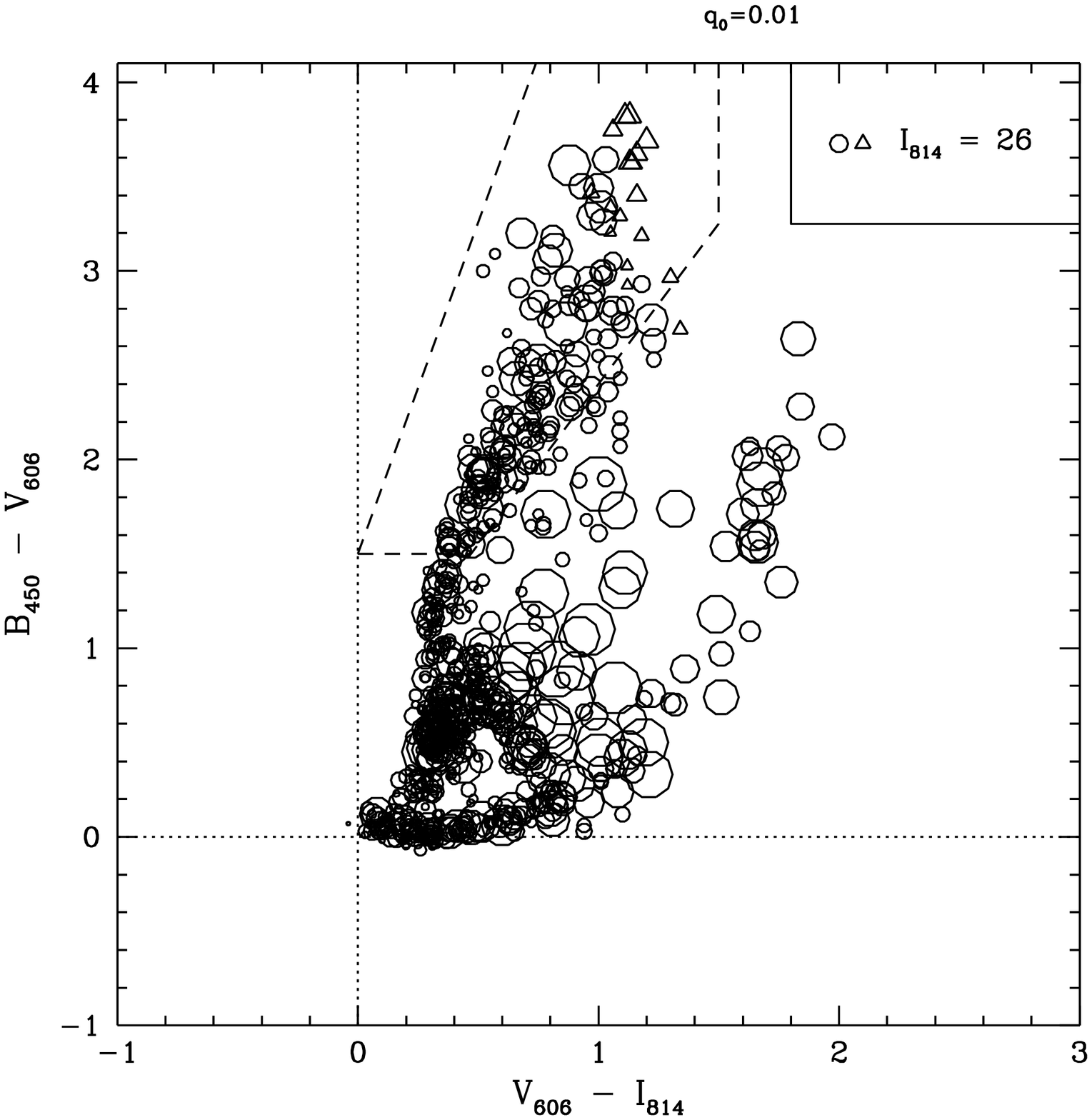}

\end{document}